\documentclass[10pt, letterpaper]{article}
\usepackage{opex3}

\usepackage{graphicx}
\usepackage{epstopdf}

\begin{document}

\title{Surface plasmons and magneto-optic activity in hexagonal Ni anti-dot arrays}

\author{Evangelos Th. Papaioannou$^{1}$, Vassilios Kapaklis$^{1}$, Emil Melander$^{1}$, Bj\"orgvin Hj\"{o}rvarsson$^{1}$, Spiridon D. Pappas$^{2}$, Piotr Patoka$^{3}$, Michael Giersig$^{3}$, Paul Fumagalli$^{3}$, Antonio Garcia-Martin$^{4}$, and Georgios Ctistis$^{5}$}

\address{$^{1}$Department of Physics and Astronomy, Uppsala University, Box 516, SE-751 20 Uppsala, Sweden}
\address{$^{2}$Engineering Science Department, School of Engineering, University of Patras, GR-26504 Patras, Greece}
\address{$^{3}$Institut f\"{u}r Experimentalphysik, Freie Universit\"{a}t Berlin, DE-14195 Berlin, Germany}
\address{$^{4}$IMM-Instituto de Microelectr\`onica de Madrid (CNM-CSIC), Isaac Newton 8, PTM, Tres Cantos, E-28760 Madrid, Spain}
\address{$^{5}$Complex Photonic Systems (COPS), MESA + Institute for Nanotechnology,University of Twente, NL-7500 AE, Enschede, The Netherlands}

\date{\today}

\begin{abstract}
The influence of surface plasmons on the magneto-optic activity in a two-dimensional hexagonal array is addressed. The experiments were performed using hexagonal array of circular holes in a ferromagnetic Ni film. Well pronounced troughs are observed in the optical reflectivity, resulting from the presence of  surface plasmons. The surface plasmons are found to strongly enhance the magneto-optic response (Kerr rotation), as compared to a continuous film of the same composition.  The influence of the hexagonal symmetry of the pattern on the coupling between the plasmonic excitations  is demonstrated, using optical diffraction measurements and theoretical calculations of the magneto-optic and of the angular dependence of the optical activity.
\end{abstract}




\section{\label{introduction}Introduction}

The use of magnetic materials in plasmonic structures allows the coupling between plasmonic and magnetic states. For example, recent results in Au/ferromagnetic layer/Au heterostructures revealed the ability to control the propagation of surface plasmon polaritons (SPPs) at the metal-ferromagnetic interface, using an applied external magnetic field \cite{Temnov,Belotelov,Torrado10}. Although the exploration of the effect on the magneto-optic (MO) activity in magneto-plasmonic structures has just started, the potential for utilization in optoelectronics \cite{Temnov}, bio-sensors \cite{Sepulveda2006}, and magnetic applications \cite{Stipe2010} are already apparent. 

The discovery of the extraordinary optical transmission through sub-wavelength hole arrays \cite{Ebbesen98} started an intensive research effort on  the optical response of plasmonic structures. The investigations of SPPs have almost exclusively been limited to metal-dielectric systems, where the metal is non-magnetic and highly conductive (Ag, Au, Al) \cite{Raether}. In the case of highly absorbing magnetic metals, such as Fe, Co, and Ni, SPPs undergo a strong internal damping, weakening the optical activity. Recently, however, plasmonic states in pure ferromagnetic patterned Ni dots were revealed \cite{Nogues2011}, while the generation of SPPs in perforated ferromagnetic metal (Fe, Co) nano-structures were demonstrated and have been used to enhance MO surface effects \cite{PhysRevB.81.054424,Ctistis2009,Phys.stat.sol.RRL}. An enhancement of MO Kerr rotation was found in hexagonal arrays of Ni nanowires embedded in an alumina matrix \cite{Diaz07} due to a surface plasmon resonance of the Ni nanowires. Ni gratings coated with Ag \cite{Newman08} also showed enhanced MO activity.  However, there are still many  unresolved questions concerning the basic principles of the underlying effects and the root of the MO enhancement arising from the generation of surface plasmons at a magnetic metal/air interface must be considered as an open scientific question. 

Here we focus on the effect of plasmonic states in an anti-dot structure of ferromagnetic material (Ni). We examine the possibility to excite SPPs on the patterned surfaces and we explore the angular dependence effects,  i.e.  the orientation of the pattern with respect to the scattering plane. The angular and energy dependent evolution of SPPs shows how the interactions of plasmonic excitation couple with the different length scales of the hexagonal pattern. Finally, we show to what extend these SPPs do influence the magneto-optic response.

\section{\label{Experimental details} Experimental and Theoretical details}

The Ni antidot arrays were prepared on Si(111) substrates, using self-assembly nano-sphere lithography with polystyrene (PS) spheres of $a = 470$ nm in diameter \cite{PhysRevB.81.054424, Ctistis2009}. After the self-assembly, the diameter of the spheres was reduced to $d = 275$ nm, using reactive ion etching. These spheres were used as shadow masks, while growing a patterned Ni, directly on the substrate. The deposition started with a 2 nm thick Ti layer, followed by a 100 nm thick Ni film, and a 2 nm thick Au layer on top. The Ti layer was used to enhance the adhesion to the substrate while the Au is used to protect the surface of the film. The self-passivation of Ni oxide prevents the degradation of the structure, which is of importance for the bare Ni surface at the rim of the holes. Ni can be regarded as a material of choice when making patterned magnetic structures, since the self-passivating oxide is very thin. The NiO layer reaches a maximum thickness of  1 nm at room temperature \cite{poulopoulos:202503, Saiki199333}.

\begin{figure}
	\begin{center}
 \includegraphics[width = 8cm]{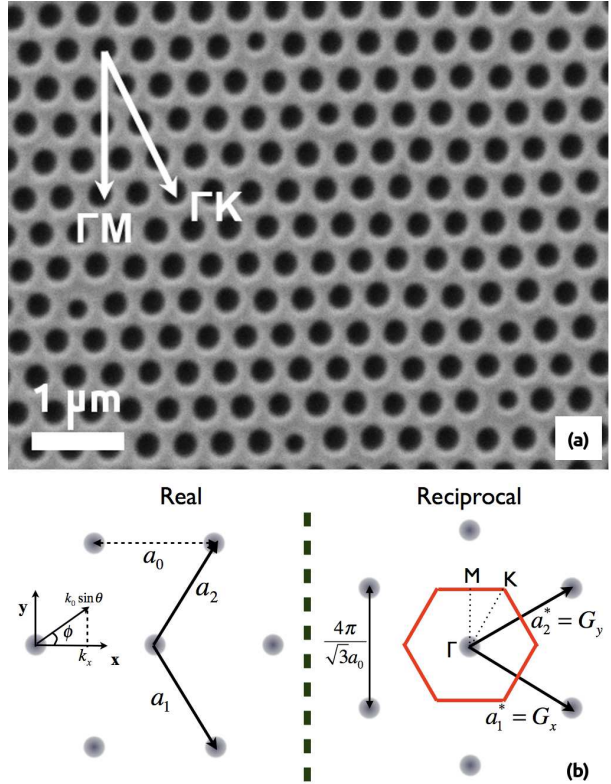}
	\caption{\label{fig:fig1} (a) Scanning electron microscopy image of the sample's surface indicating the overall good quality of the lithographic process. The sample has an average pitch size of $a = 470$ nm and hole diameter of $d = 275$ nm. (b) Real and reciprocal space of a two dimensional (2D) hexagonal lattice and the corresponding angle definitions that are used in this experiment.}
	\label{sem}
	\end{center}
\end{figure}

After the deposition, the PS mask was dissolved, resulting in a hexagonal anti-dot structure as illustrated in Fig. \ref{sem}. The ratio of the radius ($d/2$)/pitch size ($a$) was determined to be {$0.30$}, while the fraction of the holes represent 31 \%  of the total surface. A continuous film of the same structure was also fabricated as a reference sample. 

Ultraviolet (UV)-visible reflectivity spectra were recorded at room temperature using a Perkin Elmer L-35 UV-visible spectrometer, in the energy (wavelength) range $1.13 - 4.96$ eV ($250 - 1100$ nm). A 50 mm integrating sphere was used for the acquisition of the reflectivity curve. The incident light had thereby an angle of 8\textdegree \space with respect to the normal of the sample surface. The magneto-optic spectra were recorded using a magneto-optic Kerr spectrometer \cite{papaioannou:023913} operating in a polar configuration. The range in recorded energies (wavelengths) was  $0.8 - 5$ eV ($248 - 1550$ nm). The Kerr spectra were obtained in the saturation state of the samples at a magnetic field of 1.64 T.

\begin{figure}
	\begin{center}
 	\includegraphics[width = 10cm]{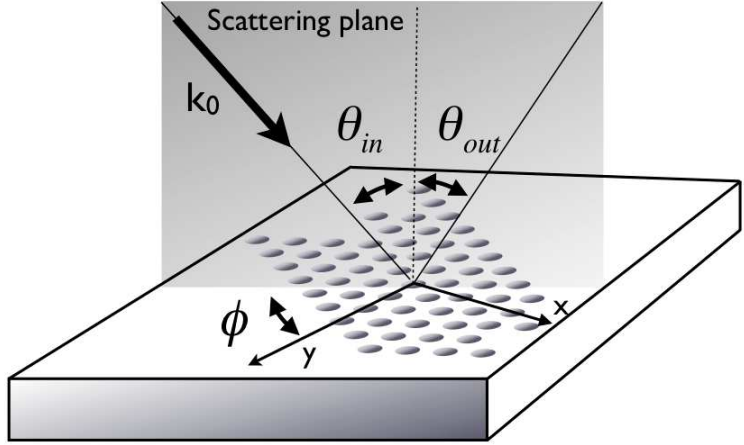}
	\caption{\label{fig:fig2} Schematic of the measurement configuration including the definition of the angles. All the measurements discussed here were performed with $\theta_{in}$=$\theta_{out}$=$\theta$. } 
	\label{scattering}
	\end{center}
\end{figure}

Angular dependent $0^{th}$-order reflectivity curves were obtained using a dedicated optical diffractometer based on a HUBER goniometer with a MC 9300 controller. This optical diffractometer consists of six motorized stages for the angular (step resolution of 1 millidegree) and positional control, monochromatic laser light sources of different wavelengths (408, 535, 660 and 787 nm), and Glan-Taylor polarizers. The scattering geometry is defined in Fig. \ref{scattering}: $\theta_{in}$ is the angle of incident light while $\theta_{out}$ is the angular position of the detector. $ \phi$ is defined as the azimuthal angle between the y-axis (defined as one of the high symmetry directions of the hexagonal lattice as shown in Fig. \ref{sem}) and the scattering plane. All the results presented here are with $\theta_{in}$=$\theta_{out}$=$\theta$, often referred to as $\theta - 2\theta$ scans.

We used a Scattering Matrix Method (SMM), specially developed to include MO effects \cite{antonio05} in periodic media, when calculating the optical and MO response of the structures. The optical and MO constants of Ni  were taken from Ref. \cite{visnov93} and those for Si from ref \cite{Palik}. The values for Au and Ti were obtained from ellipsometry measurements of 20 nm thick films. 

\section{\label{Results and Discussion}Results and Discussion}

\begin{figure}
\begin{center}
 \includegraphics[width = 8cm]{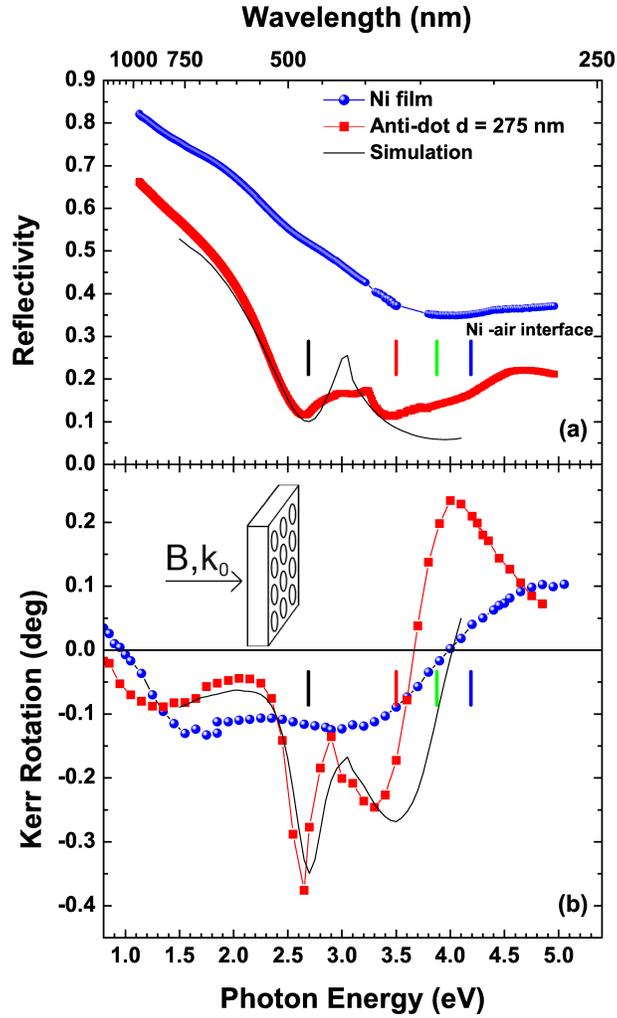}
\caption{\label{fig:fig3} (a) Reflectivity spectra as a function of wavelength of the Ni reference film and the Ni anti-dot array ($a = 470$ nm, $d = 275$ nm) at 8\textdegree\space angle of incidence. The vertical lines denote the calculated positions of reflection minima according to the Eq. \ref{eq:normalincidence} at normal incidence for a Ni film-air interface. (b) Kerr spectra of the two samples. Enhancement of the Kerr rotation is observed at positions where characteristic troughs in reflectivity occur due to SPPs excitation. The continuous lines represent the calculated reflectivity and polar Kerr rotation versus photon energy.}
\end{center}
\end{figure}

Fig. \ref{fig:fig3} (a) shows the reflectivity spectra as a function of photon energy for the hexagonal anti-dot array sample together with the continuous Ni film. As shown in the figure, the overall reflectivity of the patterned sample is lower than that of the continuous film.  Pronounced minima are observed in the reflectivity from the patterned sample, the most distinctive one at a photon energy (wavelength) $\hbar \omega = 2.65$ eV ($\lambda= 467$ nm) and a broader one from around $ \hbar \omega = 3.2$~eV ($\lambda= 387$ nm) to $\hbar \omega  = 4.4$ eV ($\lambda= 282$ nm). These are signatures of SPPs excitations, which are directly connected to the symmetry of the lattice and the optical constants of Ni, as discussed below.\\

SPPs can be resonantly excited by the coupling between free surface charges (electrons) of the metal and the incident electromagnetic field. The real part of the dispersion relation of SPPs propagating at the interface with a frequency-dependent SPP wave-vector, $k_{\mathrm{SPP}}$  can be obtained from Maxwell's equations, using the appropriate boundary conditions  \cite{barnes03},

\begin{equation}
\label{eq:dispersion}
 \mid\overrightarrow{\bf k}_{\mathrm{spp}}\mid = k_{0}\sqrt{\frac{\epsilon_{m}\epsilon_{d}}{\epsilon_{m}+\epsilon_{d}}}
\end{equation} 

\noindent
where $\epsilon_{d}$ is the frequency-dependent permittivity of the dielectric material and $\epsilon_{m}$ the real part of the frequency-dependent permittivity of the metal. These must have opposite signs, if SPPs are to be supported at the interface, condition that is satisfied for metals. For Ni at $660$ nm (wavelength that is most used in this work), the SPP wavevector for the Ni - air interface assuming optical constants of Ni \cite{Palik} ($\epsilon = -10.9615 + i 15.7966$), is found to be $k_{\mathrm{SPP}} = 1.04899 \times k_{0}$ using Eq. \ref{eq:dispersion}.

Coupling photons of a given energy with a grating structure (here the $2-$dimensional (2D) hexagonal periodic array) that provides some additional momentum, it is possible to excite surface plasmons according to the dispersion relation \cite{Raether}:

\begin{equation}
\label{eq:matching}
 \overrightarrow{\bf k}_{spp} = \overrightarrow{\bf k}_{x} \pm  i \overrightarrow {\bf G}_{x} \pm j \overrightarrow {\bf G}_{y}
\end{equation}

\noindent

where $k_{spp}$ is the surface plasmon polariton wave vector, $k_{x}$ = $ k_{0}sin \theta$ is the component of the incident wave vector that lies in the plane of the sample, $ \overrightarrow {\bf G}_{x}$, and $\overrightarrow {\bf G}_{y}$ are the basis vectors of the reciprocal hexagonal lattice and i, j are integers. 
The in-plane wave vector  $k_{x}$ of the incident light is a function of the incident angle $ \theta$ and the azimuthal angle $\phi$ between the plane of incidence and a specific lattice orientation (see Fig. \ref{fig:fig1},  \ref{fig:fig2}) and is given by:

\begin{equation}
\label{eq:wavevector}
 \overrightarrow{\bf k}_{x} = |\overrightarrow{\bf k}_{0}| sin( \theta)cos(\phi)\overrightarrow{\bf x}  + |\overrightarrow{\bf k}_{0}| sin( \theta)sin( \phi) \overrightarrow{\bf y}
\end{equation}

\noindent
  
The reciprocal vectors $ \overrightarrow {\bf G}_{x}$, and $\overrightarrow {\bf G}_{y}$ (Fig. \ref{fig:fig1} inset) are inversely proportional to the pitch size $a_{0}$, with a magnitude of $ \mid \overrightarrow {\bf G}_{x} \mid$ = $ \mid \overrightarrow {\bf G}_{y} \mid$ = $\frac{4\pi}{\sqrt{3}a_{0}}$. The reciprocal lattice is rotated $30 $\textdegree \space with respect to the real lattice as it is shown in Fig. \ref{fig:fig1} where also the corresponding angles used in the experiment are presented. The integers i, j specify the orders of the SPPs resonances.
For a hexagonal lattice, when the component of the incident wave vector $k_{x}$ is along  $ \overrightarrow {\bf G}_{x}$, the conservation of energy in Eq. \ref{eq:matching} can be written as:

\begin{eqnarray}
\label{eq:matchinghexagonal}
 \mid\overrightarrow{\bf k}_{spp}\mid &=& \biggl[ \left(\frac{2\pi}{\lambda}\sin\theta + i \frac{4\pi}{\sqrt{3}a_{0}}\right)^{2} \\\nonumber  &+&  \left(j \frac{4\pi}{\sqrt{3}a_{0}}\right)^{2} \\\nonumber
&+& \left(\frac{2\pi}{\lambda}\sin\theta +  i \frac{4\pi}{\sqrt{3}a_{0}}\right)\left(j \frac{4\pi}{\sqrt{3}a_{0}}\right)\biggr]^{1/2}
\end{eqnarray}

\noindent

Combining Eq.(\ref{eq:dispersion}) and (\ref{eq:matchinghexagonal}) we find that the angular dependent dispersion relation for an hexagonal lattice along the $ \overrightarrow {\bf G}_{x}$  direction is given by :

\begin{eqnarray}
\label{eq:angulardependent}
 \frac{\epsilon_{m}\epsilon_{d}}{\epsilon_{m}+\epsilon_{d}}&=& \sin^{2}\theta + \frac{2}{\sqrt{3}}\frac{\lambda}{a_{0}}(2i+j)\sin\theta \\\nonumber &+&   \frac{4}{3}\frac{\lambda^{2}}{a_{0}^{2}} (i^{2}+j^{2}+ij)\\\nonumber
\end{eqnarray}

\noindent

Similar calculations can be performed when the component of the incident wave vector, $k_{x} = k_{0}\sin\theta$ is along the  $ \overrightarrow {\bf G}_{x}$+ $ \overrightarrow {\bf G}_{y}$  direction.

It is worth to notice that Eq. \ref{eq:angulardependent} is derived based only on geometrical considerations and does not take into account the presence and the size of the holes and the subsequent losses and interference effects. However, it serves as  a guideline for the expected energy and angular dependence of SPPs. 
For the case of normal incidence and from Eq. \ref{eq:angulardependent} the SPPs resonances are expected at wavelengths \cite{ebbesen07}:

\begin{equation}
\label{eq:normalincidence}
 \lambda = \frac{a_{0}}{\sqrt{\frac{4}{3}(i^{2}+j^{2}+ij)}}\sqrt{\frac{\epsilon_{m}\epsilon_{d}}{\epsilon_{m}+\epsilon_{d}}}
\end{equation} 

We can now calculate the expected positions of the reflectivity minima for an angle of incidence of 0\textdegree\space, using Eq. \ref{eq:normalincidence}. 
The expected positions of the minima from a Ni hexagonal pattern are indicated as vertical lines in Fig. \ref{fig:fig3}. As seen in the Figure, the positions of the two first minima are captured, while the expected higher order minima are not clearly captured in the experimental data. Furthermore, these calculations do not provide any quantitative information on the shape of the reflectivity. Therefore, we performed quantitative calculation of the reflectivity, using the Scattering Matrix Method (SMM)\cite{antonio05}. The results are presented as a continuous line in Fig. \ref{fig:fig3} (a). The simulated curve contains the main experimental features of the reflectivity: two characteristic troughs in the energy range of   $ \hbar \omega = 1.5-4.1$ eV ($\lambda = 302- 828$ nm).  The second, broader, dip consists of three unresolved plasmonic resonances (see Eq. \ref{eq:normalincidence}) which are not resolved due to the width of the resonances, giving rise to an overlap. 

The energy dependence of the polar Kerr rotation of the patterned sample is illustrated in Fig. \ref{fig:fig3}(b), together with the results from the 100 nm thick reference layer (continuous Ni film). As expected, the reference spectrum corresponds to a typical MO spectrum of a Ni film \cite{papaioannou:023913,Zvezdin} with respect to the shape and the value of the Kerr rotation. Two distinct minima at 1.6 eV and at 3.0 eV can be identified. It also shows a change of sign, near the photon energies of 1 and 4 eV. Ni has the lowest MO activity compared to Co and Fe, corresponding roughly to the relative magnetization of these materials. Ab-initio and numerical calculations \cite{PhysRevB.45.10924} revealed a dependence of the polar Kerr rotation on the inter-band transitions (important for energies smaller than $1-2$ eV) in combination with the spin-orbit interaction and spin-orbit coupling strength \cite{Zvezdin}. However, the differences in the MO activity of Ni, as compared to Co and Fe, are not fully understood microscopically \cite{PhysRevB.45.10924}. 

The spectrum for the patterned sample shows a completely different behavior, while certain aspect of the reflectivity curve can be identified (see Fig. \ref{fig:fig3}(a)).  A sharp trough occurs around $\hbar \omega ~(\lambda) =2.7$ ($460$) eV  (nm), which coincides with a minimum in the reflectivity curve at  $\hbar \omega ~(\lambda)=2.63~(470)$ eV~(nm). At this energy, the Kerr rotation is four times larger than obtained from a continuous film. This enhancement takes place at an energy defined by the periodicity of the structure (here $a_{0} = 470$ nm) \cite{PhysRevB.81.054424}. A second minimum is obtained around $3.35~(370)$eV(nm) which corresponds to the second trough in reflectivity at the calculated resonance of $\hbar \omega ~(\lambda)= 3.5~(354)$ eV(nm). The Kerr rotation at this energy is 2.5 times bigger than obtained from the continuous film. A change in sign is observed around 3.8 eV, which resembles the changes obtained from the continuous layer, while the size of the Kerr rotation is almost doubled at the photon energy of $4.0~(310)$ eV (nm). This enhancement is also correlated to the broad minimum in the reflectivity and the two calculated plasmonic resonances after $3.5~(354)$ eV(nm).  At higher energies the MO activity decreases, approaching the one for the continuous film.  The strong enhancement for the MO signal is a result of the creation of SPPs resonances along different crystallographic orientations of the hexagonal pattern. It is known that at frequencies near the bulk plasma resonance the reflectivity shows a minimum and the MO-Kerr signal is enhanced \cite{Fuma95}. However in this case the SPPs represent another type of electronic resonances that propagates along the surface and influence the MO response.

In particular, we attribute the magneto-optic enhancement to the interplay between optical and magneto-optical effects. The magneto-optical Kerr effect has its origin in intrinsic magneto-optically active electronic transitions and is influenced by optical enhancement effects. The reduction of the reflectivity through the excitation of surface plasmons gives rise to such an optical enhancement. Simultaneously the surface plasmons intensify the electromagnetic field sensed by the Ni film. This results in an increase of the polarization conversion (or p-s light conversion) due to the magnetization of the material and to the presence of the magnetic field. The appearance of these two factors simultaneously (reduced reflectivity and enhanced polarization conversion) leads to the enhanced Kerr rotation. The aforementioned mechanism applies in several types of metal-ferromagnet-metal structures (look for example Ref. \cite{Newman08, PhysRevLett.98.077401} and references therein).

The calculated MO spectrum of the patterned sample is shown as a continuous line in Fig. \ref{fig:fig3} (b). The calculations are based on the  SMM code, specially developed to be able to include MO effects. These calculation confirms and strengthen the obtained correspondence between the changes in reflectivity and MO activity. The qualitative and quantitive agreement between theory and experiment clearly demonstrate the possibility to control and tailor the plasmonic and magneto-plasmonic resonances. 
\begin{figure}
\begin{center}
 \includegraphics[width = 8cm]{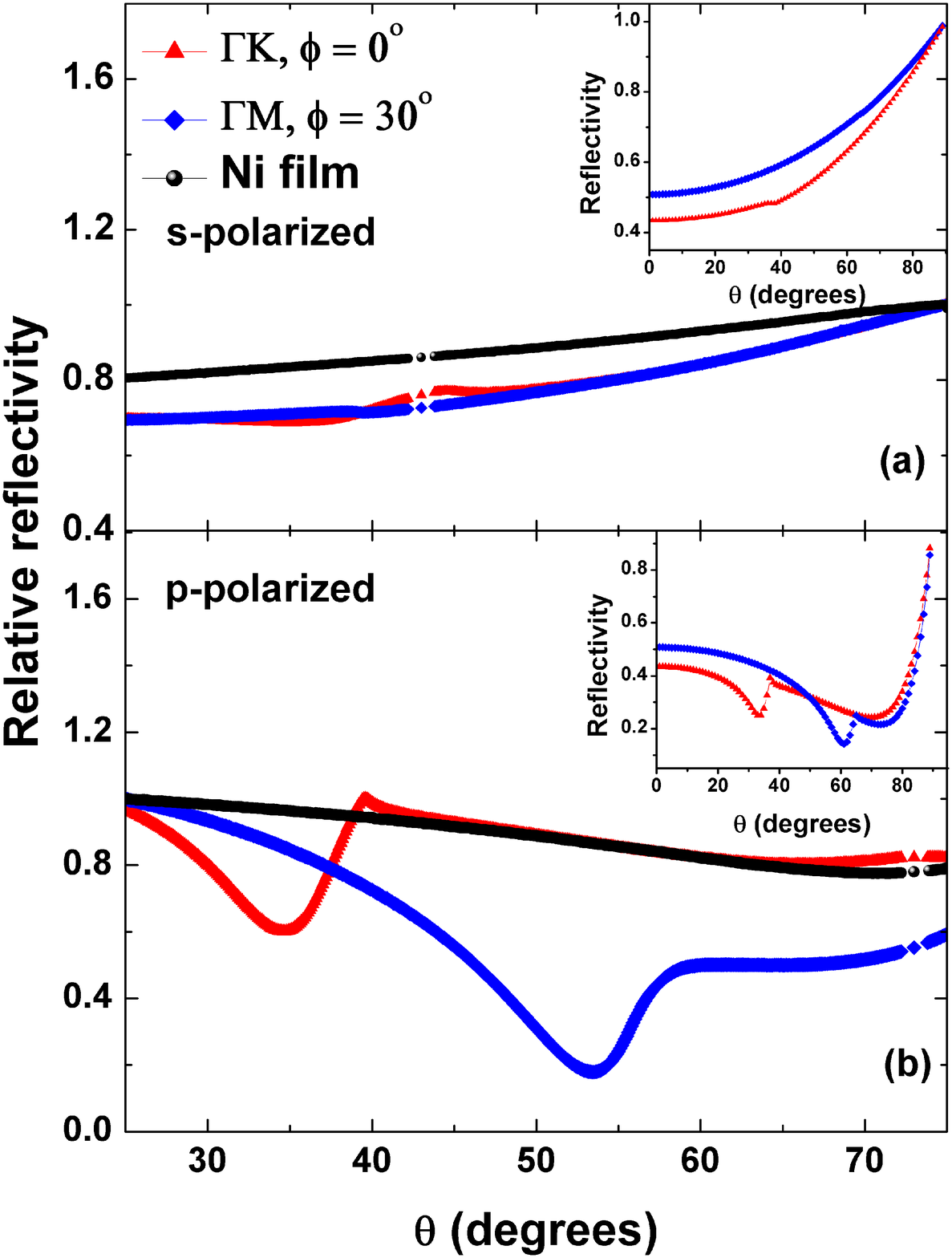}
\caption{\label{fig:fig4} Reflectivity spectra as a function of  angle of incidence for the anti-dot Ni film with $a = 470$ nm and $d = 275$ nm. Measurements performed with $\lambda = 660 $ nm for (a) s- and (b) p-polarized light. The minimum indicates the energy absorption due to SPPs excitation. The data have been normalized to their maximum value. Inset: Corresponding theoretical calculations for the experimental cases. The reflectivity is calculated relative to a reference sample. The theoretical model supports the experimental data.}
\end{center}
\end{figure}
To demonstrate the possibility to manipulate the MO activity of these samples through the SPP, we now discuss the angular dependence of the reflectivity  for $s-$ and $p-$ incident polarizations, in the two main crystallographic directions of the 2D anti-dot pattern ($\phi= 0$\textdegree\space, $30$\textdegree\space). 

The specular reflectivity curves of the continuous and patterned Ni structures are presented in Fig. \ref{fig:fig4}. The incident light is either $s-$ and $p-$ polarized with a wavelength of $\lambda = 660$nm. At a single metal-dielectric interface, no SPPs can be created for the case of $s-$ incident radiation (E vector has only one component, E$_{y}$, tangential to the interface) \cite{Maier07}. No significant change is therefore observed when changing the angle of incidence $\theta$. However, a small feature appears in the direction $\phi= 0 $\textdegree. A  $p-$ component from the incident $s-$polarization, reflection coefficient $r^{sp}$, can be induced magneto-optically since Ni is ferromagnetic and can give rise to the small resonant feature which is seen in Fig. \ref{fig:fig4} (a).  The results for $p-$ polarized light are shown in Fig. \ref{fig:fig4} (b). A remarkable drop in reflectivity around $\theta= 35 $\textdegree \space  is generated on the ferromagnetic anti-dot array for $\phi= 0 $\textdegree. The trough demonstrates the excitation of SPPs. By varying the crystallographic direction $\phi$ of the 2D antidot pattern we can tune the SPPs resonances by changing the incident angle $\theta$.
The angle $\phi= 0$\textdegree \space corresponds to the high symmetry direction $\Gamma K$  in the reciprocal space (see Fig. \ref{fig:fig1},\ref{fig:fig2}). The angle of 34.9\textdegree \space agrees well with the calculated value from Eq. \ref{eq:angulardependent}, for the case of  the (i,j)=(1,0) plasmonic mode at a Ni/air interface. Ni optical constants were taken for the corresponding wavelength $\lambda = 660$ nm  \cite{Palik}.
\begin{figure}
\begin{center}
 \includegraphics[width = 9cm]{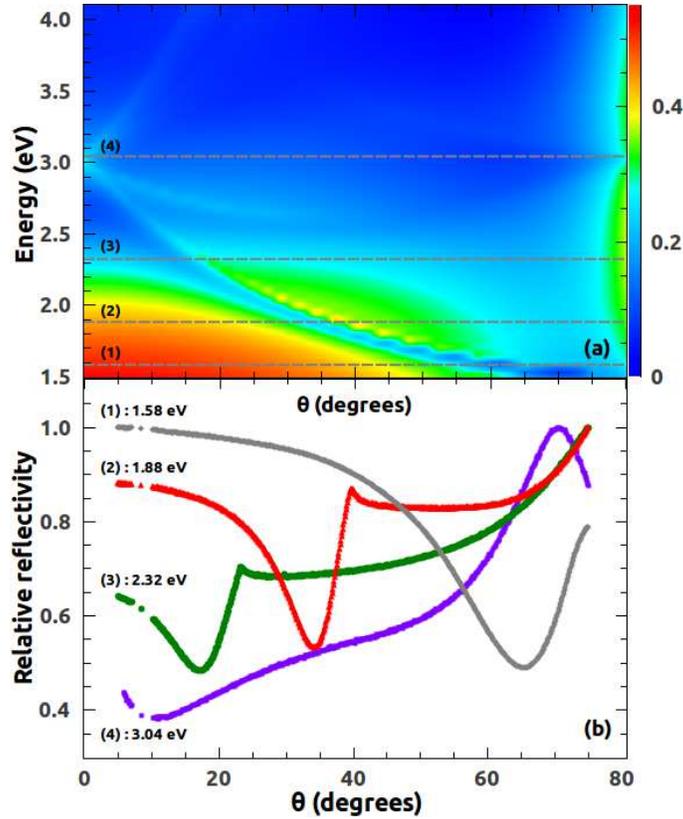}
\caption{\label{fig:fig5} (a)  Theoretical energy and angular dependent evolution of reflectivity depicted in a 2D map.  The calculations have been performed for p-polarized light at $\phi = 0$\textdegree. (b) Reflectivity of the zero order reflection for p-polarized light at $\phi = 0$\textdegree \space for four different wavelengths. The reflectivity value of each sample is normalized to their maximum value.}
\end{center}
\end{figure}
For $\phi= 30 $\textdegree \space ($\Gamma M$ direction) the resonance moves to $\theta= 53.4 $\textdegree\space with significantly larger decrease in the reflectivity. 
This behavior can be understood, when considering the length-scales probed when using these scattering geometries. At  $\phi= 0$\textdegree\space, the real space spacing is smaller than the one corresponding to  $\phi= 30$\textdegree \space (see Fig. \ref{sem}). 

The resonances will therefore appear at smaller $\theta$ when  $\phi= 0$\textdegree \space, as compared to $\phi= 30$\textdegree\space (notice the difference in the optical and x-ray reflectivity conventions). Consequently, the interplay between symmetry of the hexagonal lattice and the plasmon excitation is demonstrated  in Fig. \ref{fig:fig4}. The coupling strength of the incoming light to the hexagonal lattice depends on the incident angle, influencing the generation of plasmon resonances.

The inset in Fig. \ref{fig:fig4} (a) and (b) contain the calculated reflectivity curves for the range 0 to 90\textdegree \space in $\theta$. The experimental and theoretical results are qualitative and partially quantitatively in agreement. The largest discrepancy is in the relative change of the reflectivity, which can be attributed to both missing footprint corrections in the experiments and the exclusion of edge effects in the calculations.  When approaching the glancing incidence ($\theta= 90$\textdegree\space) the role of the hole side-walls becomes significant, which will reduce the measured reflectivity. 

Full calculation of the angular and energy dependent reflectivity ($\Gamma K$ direction of the reciprocal lattice) corresponding to $\phi= 0 $\textdegree,  is presented in Fig. \ref{fig:fig5} (a). The angular dependence of the minima in reflectivity (when varying the incident wave vector $k_{0}$), yields a shift of the minimum, to higher $\theta$ angles as the energy is decreased. The experimental results are displayed in Fig. \ref{fig:fig5} (b). These results were obtained at a constant energy, marked with dotted lines in Fig. \ref{fig:fig5} (a). The agreement between the experimental and calculated results is good, both with respect to the expected position of the minimum as well as the overall changes in the reflectivity. At 1.88 and 2.32 eV, we notice a local maxima in the intensity on the high angle side after the minima of reflectivity, which are identified as Wood's anomalies \cite{Wood35}. 

The magnetic field is expected to modify the surface plasmon state in similar structures composed of metallic films \cite{PhysRevB.77.205113} where the applied magnetic field is a second order term in the off diagonal elements of the electric permitivity tensor. Whether similar behaviour is present in our case, where we have a  magnetic metal with a smaller corresponding permitivity element (for the case of Ni: $\epsilon_{xy}(660 nm)= -0.20166391 + i 0.00708567$), needs further future investigations.

\section{\label{conclusions}Conclusions}

The optical and MO response of a hexagonal anti-dot pattern, with a pitch distance of 470 nm and hole diameter $d = 275$, are revealed to exhibit strong coupling between the SPPs and the MO activity. The intensity minima in the reflectivity coincides with the excitation of SPPs, arising from the coupling between the incident light and the 2D hexagonal lattice. The spectral position of the minima depends on the periodicity of the structures and subsequently on the different diffraction orders of the 2D grating. Large changes in the MO signal are observed at the corresponding minima in the reflectivity, confirming the importance of SPPs on the magnetic response. Angular  and energy dependent measurements of the reflectivity reveal the importance of surface plasmons and the possibility to control the coupling of SPPs in the magnetic metal by changing the scattering geometry. The use of theoretical simulations for optimizing the design and resulting properties is shown to be conceivable, which can be of importance for developing new ways to use combined magneto-optic and surface plasmon effects for manipulation of light.

\section{Acknowledgments}

Financial support from Swedish Foundation for International Cooperation in Research and Higher Education (STINT) is gratefully acknowledged. A.G.-M. acknowledges funding from  EU (NMP3-SL-2008-214107-Nanomagma), the Spanish MICINN (`MAGPLAS' MAT2008-06765-C02-01/NAN and `FUNCOAT' CONSOLIDER INGENIO 2010 CSD2008-00023), and the Comunidad de Madrid (MICROSERES-CM' S2009/TIC-1476). GC acknowledges financial support by NWO-nano.


\end{document}